\documentclass[aps,pre,preprint,groupedaddress,showpacs]{revtex4} 
\usepackage{graphicx}
\usepackage{bm}
\usepackage{setspace}                         
\usepackage{amsmath}
\usepackage{longtable}
\newcommand{\mth}{$m_{th}$}

\begin{document}
\title{Network of Recurrent events - A case study of Japan}
\author{Revathi, P. G.}
\email{revathi.pg@gmail.com}
\author{Krishna Mohan, T. R.}
\email{kmohan@cmmacs.ernet.in}

\date{\today}

\begin{abstract}
A recently proposed method of constructing seismic networks from ``record breaking events'' from the earthquake catalog of California (Phy. Rev. E, 77 {\bf 6},066104, 2008) was successfull in establishing causal features to seismicity and arrive at estimates for rupture length and its scaling with magnitude. The results of our implementation of this procedure on the earthquake catalog of Japan establishes the robustness of the procedure. Additionally, we  find that the temporal distributions are able to detect heterogeneties  in the seismicity of the region.  
\end{abstract}
\maketitle
\section{Introduction}
Earthquakes are a prime example of a natural phenomenon-- where the spatial and temporal coordinates of an event can be measured with a fair degree of accuracy even as the dynamics of the process is not fully understood. The unpredictability  of an earthquake occurence has  not been reduced by more numerous or more accurate measurements. The difficulties of seismic analysis  involves understanding the roles of the static/geometrical heterogeneities in fault zones, the dynamic heterogeneities that result as stress builds up in the moving lithospheric plates and the possible interactions between the two~\cite{BS2004}.  The multidimensional character of seismicity makes it difficult to model it and  this  is further hampered by the restrictions of having to apply our understanding of rock friction from lab based results  to the enormous length and time scales associated with earthquakes. In order to develop a realistic model for seismicity and for assessing the seismic potential of a region,  the factors mentioned above needs to be estimated correctly. This will involve being able to measure the stress and strain at all points along the active fault - which is currently impossible ~\cite{Jdetal2006}.

The observed clustering of earthquakes  suggests that events are correlated in space-time. Thus, studies of earthquake correlations are  important for  understanding the dynamics of the process and for evolving prediction algorithms. In recent times, a large body of work in seismic studies involves relating the topology of  complex networks constructed from earthquake catalogs of a region  to the spatio-temporal patterns exhibited by  seismicity there. Such an approach is based on the idea that the patterns of seismicity may shed light on the fundamental physics, since these patterns are the emergent processes of the underlying many body nonlinear system\cite{Jdetal2006}. New patterns in the clustering of seismicity in space and time, new parameters that characterise the seismicity in a region, causal connections between subsequent events, novel methods to estimate the rupture length and its scaling with magnitude -- are some of the prominent outcomes of these catalog based studies ( see for eg. \cite{AbeSuz2004,BaPcz2004, BaPcz2005, Eisneretal2005, AbeSuz2006, AbeSuz2006-2, Ba2006, Grassetal2007}). The central idea contained in  these studies is to treat all events in the catalog on the same footing. That is, the space - time window based distinctions of events as foreshocks - mainshocks - aftershocks is avoided. Causal connections are extended to beyond immediately subsequent events and also an event can have more than one correlated predecessor. In these studies, correlations between events have been  established from different perspectives and complex networks are constructed by linking correlated events.  For example, the region studied may be divided into small cubic cells. Each cell becomes a node in the network if an event happens within it. The network is  linked by directed edges between nodes representing  successive events in the catalog~\cite{AbeSuz2004, AbeSuz2006, AbeSuz2006-2}. Alternately, correlations have been quantified with a metric which incorporates the fractal dimension $d_{f}$ and the $b$ value of the catalog used. Links are made between highly correlated events with directed edges~\cite{BaPcz2004,BaPcz2005,Ba2006}.

 A third  procedure involves identifying {\em record breaking recurrences} of an event  and linking each event in the catalog to its recurrences with directed edges~\cite{Jdetal2006,Grassetal2007}. This  procedure,  illustrated  on the earthquake catalog of Southern California  had evolved a general method for inferring causal connections. The patterns exhibited by a network constructed from the actual catalog of the region was compared with the network of a randomly shuffled catalog (acausal) in order to establish causal features. In addition, this approach provides a method of estimating the rupture length,  its scaling with magnitude and in identifying  new scaling laws associated with the seismicity of the region. 

In this paper, we apply the procedure developed by the authors\cite{Jdetal2006, Grassetal2007} on the earthquake catalog of Japan, in order to study the patterns exhibited by its  network  and to examine the robustness of the procedure for different  seismic regions. This paper is organised as follows:-  Section II  briefly describes  the construction of the network, Section III will describe the catalogs used for the analysis,  Section IV contains the results and a discussion of it and in Section V we conclude.
\section{Construction of the network}
The procedure described in \cite{Grassetal2007} is followed here to identify the recurrences and to construct the network. A directed network is constructed by linking each event $i$, (i = 1,2,3,$\cdots$\ ,N)
 in a catalog of $N$ events with a magnitude threshold of $m_{th}$,  with an edge pointing from it ( main event) to its recurrence. Each recurrence $j$,  ( j = i+1,$\cdots$\ ,N) is a {\em record breaking event} which is  closer to the main event $i$ than any event $k$ in the intervening period $t_{j} - t_{i}$ where $t_{j} > t_{k} > t_{i}$. This implies that the $(i+1)^{th}$ event  is the first recurrence of the $i^{th}$ event for $i = 1,2,\cdots,N-1$  and is always a recurrence however distant it might be. In this way, each event gets to have its own list of recurrences which follow it in time. This definition of a recurrence is based solely on the spatiotemporal relations between events in the catalog and has no arbitrariness in its selection. Also, any event identified as a recurrence of a main event does not loose its status if the period of the catalog is enlarged later, as this will only result in the list of recurrences getting longer if the added  events satisfy the condition of coming closer to the main event in the intervening period. The network is constructed by considering every event in the catalog as a node and each recurrence is represented by a directed edge between the pair of events, directed according to the time ordering of the earthquakes. The two variables associated with each edge  are -- $l_{ij}$, the distance between the epicentres of the two events and $t_{ij}$, the time interval between the them. Each node has an out-degree equal to its number of  outward edges (the number of its recurrences) and an in-degree equal to  the number of events it is counted to be a recurrence of.  The first event in the catalog will not have any incoming edges to it and the last event in the catalog will have no outgoing edges from it.  Note that the recurrences defined here do not necessarily relate to the conventional methods of defining aftershocks and thus the largest magnitude event in the network need not have the most number of recurrences. The overall structure of the network represented by the outgoing links and the incoming links describes the clustering of seismicity in the region studied.
\section{The Catalog}
The catalog studied here is the  Japan  University Network Earthquake Catalog~\cite{jap_catalog} covering the  rectangular region  $(126.433^{\circ}\mathrm{E}$--$148.0^{\circ}\mathrm{E})$ longitude and $(25.730^{\circ}\mathrm{N}$--$47.831^{\circ}\mathrm{N})$ latitude, for the period  between January, 1986--December, 1998. The catalog has a threshold magnitude of $m_{JMA}$ =$2.0$. However, we have limited our study to events with magnitude ~$\geq 2.5$. The total number of such events are  $114804$. In order to verify the dependance of the network properties on magnitude threshold and on the number of events, subsets with higher magnitude thresholds {\it viz ~ $3.0$, $3.5$, $4.0$ and $4.5$}  which contain $53525$, $20067$, $6268$ and $2038$ events respectively and a subcatalog with \mth\  = 3.0 for a shorter period from 1986 --1992 with 22975 events, were generated from the main catalog. The $b$ value = $-0.87$ was  obtained from a  plot of $\log N(m)$ {\it vs} $m$, where $N(m)$ is the cumulative number of earthquakes of magnitudes $\geq m$,  and  the linear fit was good in the range $2.5 \leq m \leq 7$. In order to distinguish the causal features of seismicity, networks  were constructed from the shuffled  catalog, where we have adopted an identical  shuffling procedure as in \cite{Grassetal2007}. Subsets and sub catalogs of the shuffled catalog are also made for the same \mth\ values as for the actual catalog. We propose to contrast  the features of their in/out-degree, recurrent length and recurrent time distributions  in order to relate  the features of the  seismic network of Japan   with that of California~\cite{Grassetal2007}.

\begin{figure}[h]
\caption{A spatial plot of the earthquake catalog~\cite{jap_catalog}  with \mth\ =2.5 for the period 1986 -1998. The catalog contains 114804 events.}
\label{spatial}
\includegraphics[angle =270, scale = 0.6]{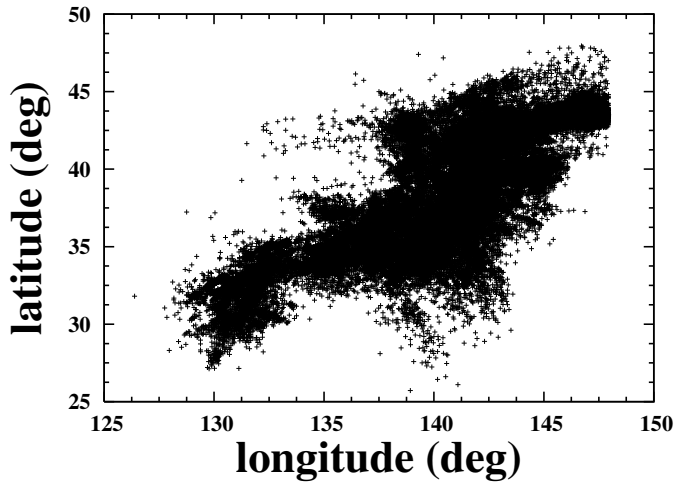}
\end{figure}

\section{Results}

\begin{figure}[h]
\caption{The panels (a) and (b) in the figure represent the degree distributions for the actual and the shuffled catalogs respectively. The points represent the out-degree and   lines of the same colour as the points  represent  the corressponding in-degree distributions for the different \mth\ values shown in the legend of the graph. The in-degree, for all \mth\ values  is almost a Poisson distribution for both the actual and the shuffled catalogs. The out-degree of the actual catalog shows significant deviations from Poisson behaviour for low and higher degree values. In case of the shuffled catalog, the marginal deviation from Poisson behaviour is only towards the higher degree values.}
\label{deg_dist}
\includegraphics[]{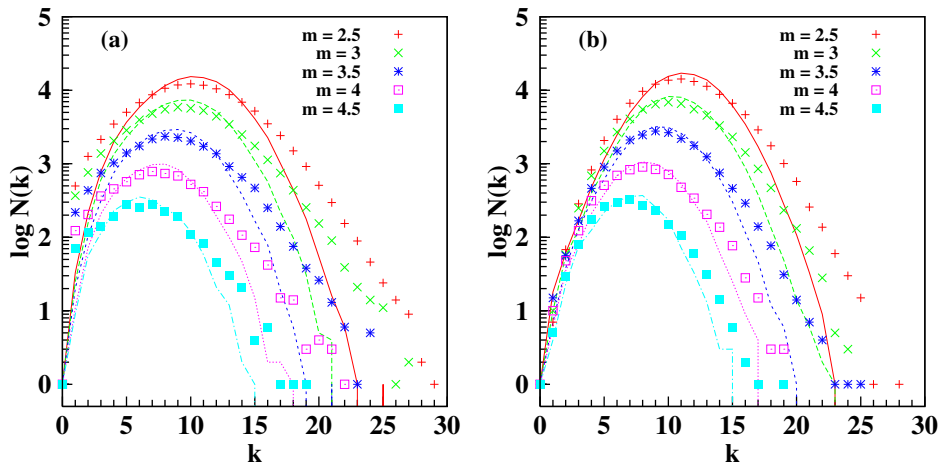}
\end{figure}

\subsection{Degree distributions}
\begin{figure}[h]
\caption{The plot shows the variation of the mean degree, $\langle k \rangle$ with the number of events $N$. The lines represent the best fits from which the slope ie estimated. The shuffled catalog, which is equivalent to a random process, shows a slope $= 1$ while the actual catalog shows a lower value of $\approx$~0.91.}
\label{meandeg}
\includegraphics[]{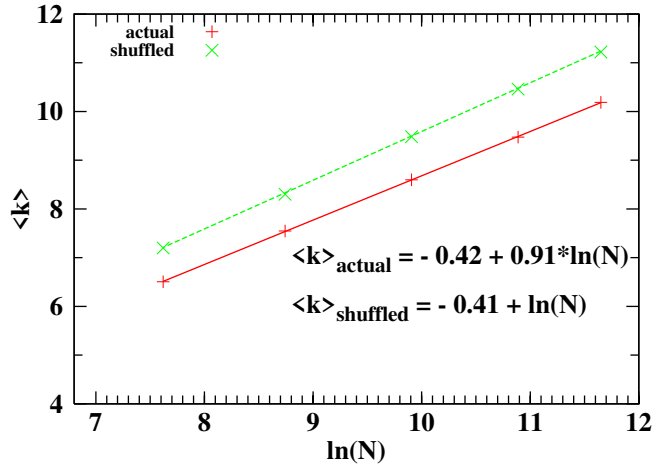}
\end{figure}

The in-degree and out-degree of every node in each of the networks of different \mth\ is estimated and the histograms in Fig.~\ref{deg_dist}(a) show the degree distributions for the actual catalog. The  points and the lines of the same colour  represent the out-degree  and the in-degree distributions respectively, for each  \mth\ value. As observed for California~\cite{Grassetal2007}, the in-degree histograms are all  Poisson in nature while the out-degrees show an  excess of  nodes with lesser and higher degree when compared to a Poisson distribution of the same mean degree. The panel $(b)$ in Fig.~\ref{deg_dist} represents the in/out-degree histograms for the shuffled catalog which is equivalent to random process. The in-degree for the shuffled catalog is observed to be  a Poisson distribution. In a random network, the out-degree distributions are also expected to be Poisson. The shuffled catalog has an out-degree  which  though not a  typical  Poisson distribution,  shows significant differences when compared to the out-degree distributions of the actual network especially for the lower degree values. For each network, the mean in-degree is equal to the mean out-degree.  As expected, this decreases with \mth\  because the number of nodes and therefore, the  number of links decrease with \mth. For a Poisson distribution, it is expected that $\langle k \rangle \approx \log N$, which is what we observe in Fig.~\ref{meandeg} for the shuffled catalog. For the actual catalog however, the number of links are  less than that of a random network and we get $\langle k \rangle \approx 0.91\log N$. The value $0.91$ is higher than that reported for California in ~\cite{Grassetal2007} and this means that larger number of links are possible for the nodes in the Japanese seismic network.

\subsection{Distance distributions}

\begin{figure}[h]
\caption{The figure shows the distribution of the distances of recurrences for the different magnitude thresholds. The panel marked $(a)$ is for the actual catalog and the one marked $(b)$  for the shuffled catalog. The black points in both the plots represent a smaller catalog(1986 -1992) with  \mth\ = 3.0. For the actual catalog, the size of the catalog does not affect the position of the peak as seen in the peak positions of the green and black points. However, for the shuffled catalog, the distance distribution varies  with size of the catalog. The inset shows the data collapse achieved by rescaling the distances and the distributions according to eqn.~\ref{eqn1}. The red line in the inset represents the pre-factor $L_{o}$ in the scaling law for the rupture length given by $ l^{\ast}(m) = L_{o} \times 10^{0.41 m_{th}}$, where $L_{o}$ = 0.066 km}
\label{RL}
\includegraphics[]{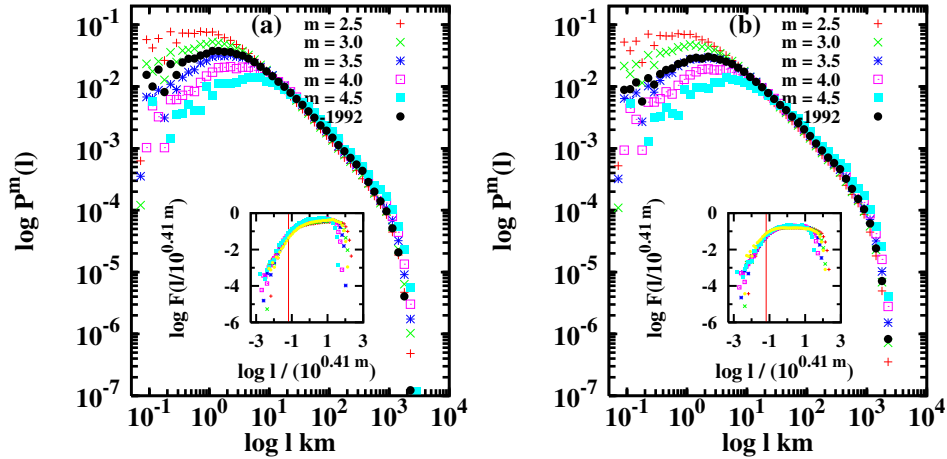}
\end{figure}

We use  the great circle distance for computing the distance \ $l_{ij}$\  between two event locations. This is calculated using the  Haversine formula~\cite{haversine}: if $(\phi_i, \lambda_i)$ and  $(\phi_j, \lambda_j)$ are the (latitude, longitude) values for the two event locations and, $\Delta \phi = \phi_j - \phi_i$ and  $\Delta \lambda = \lambda_i - \lambda_j$, then \[ l_{ij} = R \Delta \sigma\] where $\Delta \sigma$, the  spherical distance, or central angle, between the two event locations is \[ \Delta \sigma = 2 \arcsin \left( \sqrt{\sin^2 \left(\frac{\Delta \phi}{2}\right) + \cos(\phi_i) \cos(\phi_j) \sin^2 \left(\frac{\Delta \lambda}{2}\right)} \right). \]
$R$, the radius of the earth,  is taken as $6367~ km$.
The PDF of the recurrent distances for all the $i-j$ recurrent pairs in the actual network is shown in a $\log-\log$ plot of $P_{m}(l)$ versus $l$ in Fig.~\ref{RL}(a). The different coloured points are for networks with different \mth values. All the plots show that the PDF is unimodal and exhibits a peak at a typical distance $l^{\ast}_{m}$. The position of the peak  is invariant with respect to the length of the catalog as can be seen from the peak positions of the green and  black points, which corresspond to different catalog sizes  having the same \mth\ value of $3.0$. Though we have not shown it in the figure, we have verified this for the other \mth\ values also. The peak position however, depends upon \mth\ and increases with it. Beyond the peak the distribution shows a power law decrease with an exponent $\approx 1.2$ upto a cut-off which represents the size of the region studied. Since the least count of the catalog for latitude and longitude values is $0.001^{o}$, which translates to a least measurable distance $\approx 0.1$km, we are unable to resolve the points at distances $<$ 100mts. This affects the unnambiguous recognition of the peak for the lower \mth\ plots. The position of the peaks for each plot ($l^{\ast}_{m}$) was estimated by fitting a quadratic near the region of the peak. The variation of $l^{\ast}_{m}$ with $m$ was studied to arrive at the rescaling parameter of $0.41$ for the inset data collapse in Fig.~\ref{RL}(a). The data collapse obeys the relation \begin{equation}  P^{m}(l) \approx l^{-1.2} F(l/10^{0.41m_{th}})\label{eqn1} \end{equation} The scaling function $F$ has two regimes-- a power law increase with an exponent $\approx 1.8$  for small arguments and a constant regime at larger arguments. The characteristic distance of the peaks can be written as \begin{equation} l^{*}(m) = L_{o}\times 10^{0.41m_{th}} \label{eqn2} \end{equation} where $L_{o} = 0.066$. The distance distributions for the shuffled catalog in Fig.~\ref{RL}(b) shows the same overall shape as for the actual catalog. We also find that the peaks shift to higher values of $l$ as \mth\ increases. However, one important variation  observed here was that the position of the peak is not independant of $N$. This  striking and important feature was reported for the California catalog too \cite{Grassetal2007}. The invariance of $l^{\ast}_{m}$ for the actual catalog  was attributed to the causal structure of seismicity and is therefore  expected to reflect the features  of the underlying dynamics \cite{Grassetal2007}. 
\subsubsection{Identifcation with rupture length}
\begin{figure}
\caption{The distribution of distances of the first recurrence for the different \mth\ values is shown here. The peaks at the characteristic distances that was observed in fig.~\ref{RL} is visible here too. The inset shows the datacollpase achieved by rescaling the distances and the distributions according to eqn.~\ref{firstrl}. }
\label{firstRL}
\includegraphics[]{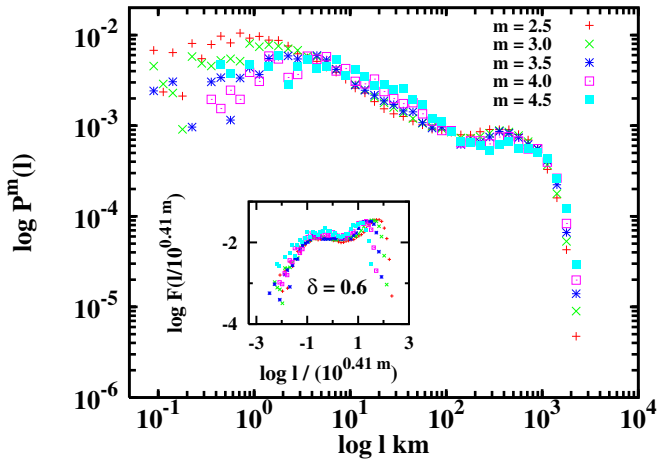}
\end{figure}

An  useful outcome of constructing such a network of recurrent events  is,  being able to arrive at an estimate for the rupture length in the region, from its earthquake catalog. The values of $l^{\ast}_{m}$ are identified as the rupture lengths and in the case of the California region the values obtained were reported to be in good agreement with the estimates given in \cite{Kagan2002,Wells1994}. The identification of $l^{\ast}_{m}$ as the rupture length is  further established by the appearance of  the same distances in the distribution of the distances of the {\em first} recurrences. The distribution of the first recurrences for California had the form \begin{equation} P_1^{m}(l) = l^{-\delta_{r}} \bar{F}(l/10^{0.45 m_{th}})\end{equation}  with $\delta_{r} = 0.6$.  The first recurrence of an event $i$ is the distance between the $i^{th}$ and $(i+1)^{th}$ event. The distances of the first recurrence is estimated for all the events in the catalog and its distribution for the actual catalog is shown in fig.\ref{firstRL}. We also observe the peaks for the different \mth\ values appearing at the same distances  as in fig.~\ref{RL} and therefore the $l^{\ast}_{m}$  values estimated from  fig.~\ref{RL} can be considered as the rupture lengths as indicated in \cite{Grassetal2007}. The distribution of first recurrences in our analysis can be written as  \begin{equation} P_1^{m}(l) = l^{-\delta_{r}} \bar{F}(l/10^{0.41 m_{th}})\label{firstrl} \end{equation} with $\delta_{r} = 0.6$ and the scaling function $\bar{F}$ has the same form as $F$ in  fig.~\ref{RL}.

The rupture lengths for Japan estimated using eqn.~\ref{eqn2} compares favourably with the values estimated using  $ L_{R}(m') \approx 0.02   \times 10^{0.5m{'}}$ ~\cite{Kagan2002} and $ L_{R}(m') \approx 0.018 \times 10^{0.46 m{'}}$ ~\cite{Wells1994} as seen from  Table~\ref{table1}

\begin{table}[h]
\caption{The values tabulated here show a comparison of the rupture lengths estimated by this procedure with  other estimates  in~\cite{Kagan2002, Wells1994}. Note that the magnitude scale used in ~\cite{Kagan2002, Wells1994} are moment magnitude ($M_{w}$). We have used the relation $M_{w} = 0.58M_{JMA} + 2.25$ ~\cite{Scord2005} in order to convert the JMA magnitude values to its  equivalent moment magnitude.}
\begin{tabular}{|l|l|l|l|l|}
\hline
$M_{JMA}$&\ $M_{w}$\ &$= 0.02\times 10^{0.5m{'}}$km&$= 0.018 \times 10^{0.46 m{'}}$km& = $0.066\times 10^{0.41m_{th}}$km\\
\hline
3.0&3.99&1.97&1.23&1.12\\
3.5&4.28&2.76&1.67&1.79\\
4.0&4.57&3.85&2.27&2.88\\
4.5&4.86&5.38&3.09&4.61\\
\hline
\end{tabular}
\label{table1}
\end{table}

\subsection{Temporal distribution of recurrences}
\begin{figure}[h]
\caption{The distributions of the recurrent times for different \mth\ values is shown here in panels $(a)$ and $(b)$ for the actual and shuffled catalogs  respectively. Significant differences can be observed between the two distributions. However,  the dependence of the distributions on $m$ in the actual catalog is  , a  major departure from what is expected from a causal network~\cite{Grassetal2007}.}
\label{RT}
\includegraphics[]{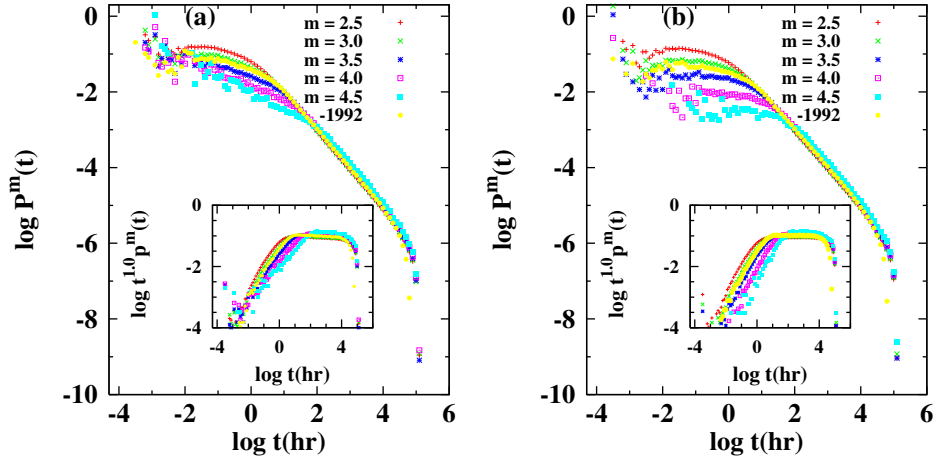}
\end{figure}
The PDF of the time intervals $t_{ij}$,  between events and their recurrences for different \mth\ values is shown in a $\log-\log$ plot in fig.~\ref{RT}(a) for the actual catalog and in the panel marked (b) for the shuffled catalog. As observed for the  California region, for intermediate times the distributions show a power law decay with an exponent $\approx 1.0$ with an observational cut-off at the longest time scales. The distributions for the different \mth\ values are all collapsed showing no dependance on magnitude in this region. However, in the region of the shorter time scales, the different distributions show a dependance on $m$. The power law exponent in this region is much less than $1.0$ and varies from $-0.3$ to $-0.45$ as \mth\ increases from $2.5$ to $4.5$. The range over which this smaller exponent holds good depends on \mth\ and increases with it. This aspect of the recurrent time distributions is at variance with that observed for California, where the temporal distributions were found to be independent of $m$ for the actual catalog and showed a single power law regime  over almost the entire range. This behaviour was attributed to the causal features of seismicity. The recurrent time distributions for the shuffled catalog in  fig.~\ref{RT}(b) shows a more prominent dependence on $m$, which  is expected for a random network~\cite{Grassetal2007}. That the actual catalog for Japan too shows characteristics attributed to an acausal network signifies a random behaviour in the temporal distributions in the case of Japan.  

\section{Conclusions}
The general procedure that was introduced to infer causal structure  from clusters of localised events was illustrated on the earthquake catalog of California. Our results from adopting the procedure on the earthquake catalog of Japan confirms its robustness. The construction of the network involves establishing correlations between nodes without making any assumptions regarding the nature of the correlations. The distinctive features of the seismic network when compared to an acausal network was attributed to the causal connections between events. Our results show that features of random networks are universal with the shuffled catalogs of both the regions showing identical network properties. The network from the actual catalog on the other hand shows some region specific features along with exhibiting typical causal attributes. The estimates of rupture length which arise as natural outcomes of the network construction are in good agreement with values estimated by other methods. The anamolies that are shown by the recurrent time distributions in the case of Japan when compared to expectations for a causal network  could be a result of the heterogenous nature of seismicity in the region. Japan experiences a high level of seismicity due to its position at the cusp of the Pacific-Philippine-Eurasian triple plate junction. It has a predominantly compressional tectonic regime as compared to Western United States which is dominantly extensional~\cite{Brodsky2006}. We expect  this aspect, along with the fact that different recurrence intervals  exist in Japan for earthquakes in subduction zones as compared to areas along crustal faults,  to influence the recurrent time distributions which is  shown in our results here. However analysis of catalogs for other such regions would have to be done in order to confirm this stand.

\bibliographystyle{unsrt}
\bibliography{synopsis}

\end{document}